# Bioinspired rational design of multi-material 3D printed soft-hard interfaces

M. C. Saldívar[a*ᵗ], E. Tay[aᵗ], E. L. Doubrovski[b], M. J. Mirzaali[a], A. A. Zadpoor[a]

*[a] Department of Biomechanical Engineering, Faculty of Mechanical, Maritime, and Materials Engineering, Delft University of Technology (TU Delft), Mekelweg 2, 2628 CD, Delft, The Netherlands*
*[b] Faculty of Industrial Design Engineering (IDE), Delft University of Technology (TU Delft), Landbergstraat, 15, 2628 CE, Delft, The Netherlands*

* Corresponding author. e-mail: m.cruzsaldivar@tudelft.nl.
ᵗ Both authors contributed equally to this study.

**ABSTRACT**

Durable interfacing of hard and soft materials is a major design challenge caused by the ensuing stress concentrations. In nature, soft-hard interfaces exhibit remarkable mechanical performance, with failures rarely happening at the interface but in the hard or soft material. This superior performance is mechanistically linked to such design features as hierarchical structures, multiple types of interlocking, and functional gradients. Here, we mimic these strategies to design efficient soft-hard interfaces using voxel-based multi-material 3D printing. We designed several types of soft-hard interfaces with interfacial functional gradients and various types of bio-inspired interlocking mechanisms. The geometrical designs were based on triply periodic minimal surfaces (*i.e.*, octo, diamond, and gyroid), collagen-like triple helices, and randomly distributed particles. The length of the gradient was varied for all cases. We utilized a combination of the finite element method and experimental techniques, including uniaxial tensile tests, quad-lap shear tests, and full-field strain measurement using digital image correlation, to characterize the mechanical performance of different groups. The designs based on gyroid minimal surfaces and those incorporating a random distribution of particles exhibited the best performance. The analysis of the best performing designs (*i.e.,* the gyroid, collagen, and particle designs) suggests that smooth interdigitated connections, compliant gradient transitions, and either decreasing or constraining the strain concentrations regions between the hard and soft phases led to simultaneously strong and tough interfaces. Increasing the gradient length was only beneficial when the resulting interface geometry reduced strain concentrations (*e.g.*, in collagen and particles). Combining the gyroid-based architecture with a random distribution of particles yielded the best-performing soft-hard interface, with strengths approaching the upper limit of the possible strengths and up to 50% toughness enhancement as compared to the control group. The stellar performance of this design is due to the presence of all the abovementioned toughening mechanisms that work synergistically to create such an effective soft-hard interface.

## 1. INTRODUCTION

Joining materials with dissimilar mechanical properties is inherently challenging due to the complexities present at the soft-hard interfaces [1–3]. These complexities include the different load-carrying capacities of both materials, interfacial damages caused by the failure of any adhesives present at the interface, and the stress concentrations caused by the sudden changes in the material properties [4–7]. Among those factors, the lattermost is particularly concerning because interfacial microarchitecture and geometry play key roles in the development of stress singularities [8]. In contrast, several millennia of evolution has endowed natural architected structures with remarkable mechanical properties that originate from their complex yet highly efficient arrangements of mechanically dissimilar phases [9–11]. Given the failure of engineered constructs in reproducing the high level of efficiency exhibited by natural materials, it is important to understand and mimic the naturally occurring design strategies.

A prime example of such high performing interface is the tendon enthesis, where the soft tendon connects to the much stiffer bony tissue along a relatively short transitional length [12], employing an efficient mixture of design features, such as morphological interdigitations and anisotropic orientations [13–16]. Moreover, functional gradients (FGs) enable a smooth transition of material properties from bone to tendon, reducing interfacial stresses [17–21]. The synergy of these mechanisms makes the bone-tendon connection highly efficient [12].

To date, a major impediment to the application of such design features has been the lack of suitable manufacturing techniques. The emergence of multi-material additive manufacturing (=3D printing) techniques has addressed this limitation and has enabled us to closely emulate the abovementioned natural design paradigms. In particular, controlling the type of the deposited material at the level of individual voxels makes polyjet multi-material 3D printing highly suitable for the emulation of natural soft-hard interfaces [22–24].

Many types of architectures could be used as a basis for the design of biomimetic interfaces. Here, we selected a few types of architectures to study the effects of architecture types and design parameters on the mechanical performance of the resulting soft-hard interfaces. Triply periodic minimal surfaces (TPMS) [25–27] were one of the selected architectures because they offer a large surface area to volume ratios and high genus values, both of which are highly beneficial for an enhanced interlocking of the interfacing phases. A high genus value means that there are multiple surface-connected yet volume-separated compartments available in the architecture of the material. Each of those compartments could be occupied by one of the interfacing phases. In this way, the phases interlock volumetrically across the vast surface area of the unit cells. A higher contact area between the material phases can also reduce strain concentrations. Architectures based on collagen-like helices [28,29] were also considered because they facilitate the creation of functional gradients while offering open cells and high surface areas. The design matrix was complemented by including randomly distributed particles, which are known to generate smooth functional gradients and arrest propagating cracks [30]. Moreover, the random distribution of particles can be integrated into multi-hierarchical arrangements [31] to prevent failure within the interface region.

We used both experiments and computational models to compare the various design options mentioned above and to elucidate the mechanisms determining the relative performance of different architectures. In particular, we studied the relation between the internal geometry, the type of the transition function, and the contact surface on the one hand and the mechanical characteristics of the soft-hard interfaces, and the ensuing strain concentrations [32] on the other. This novel approach provides us with a pathway towards a better understanding of the mechanisms at play in the design of soft-hard interfaces and enables us to devise some design guidelines for improving the mechanical performance of bioinspired soft-hard interfaces with potential applications in tissue engineering, soft robotics, and architected materials.

## 2. MATERIALS AND METHODS

### 2.1. 3D printing setup

We used a poly-jet multi-material 3D printer (ObjetJ735 Connex3, Stratasys® Ltd., USA) with voxel-level control to manufacture our biomimetic soft-hard interfaces. The commercially available photopolymers VeroCyan™ (RGD841, Stratasys® Ltd., USA) and Agilus30™ Clear (FLX935, Stratasys® Ltd., USA) were used for the hard and soft phases, respectively. Stacks of binary images detailing the type of the deposited material at each voxel were provided as input to the printer. In this approach, every image represents a layer of the 3D design. The maximum printing resolution was 300×600×900 dpi. We used the minimum edge size to print cube-shaped voxels with an edge length of 85 μm.

### 2.2. Design and manufacturing of tensile test interfaces

We considered the narrow section of a standard tensile test specimen shape (type IV) described in ASTM D638-14 [33] to constrain the dimensions of our designs (Figure 1A). These dimensions allowed for a design region of 384×96×48 voxels (32.51×8.1×4.0 mm$^3$). Within these regions, we kept the length of the soft region constant ($W_S = 8.128$ mm) and varied two interface parameters, namely the width ($W_G$) and geometrical design of the functional gradient (Figure 1B). We selected three different values of $W_G$ (*i.e.,* 4.064, 8.128, 12.192 mm, equivalent to 48, 96, and 144 voxels), wherein we linearly varied the volume fraction of the hard phase ($\rho$) from 0 to 100%. We discretized these functions using cubic-shaped unit-cells with five different geometries, including three TPMS-based architectures (*i.e.*, octo (OC), diamond (DI), and gyroid (GY)), biomimetic collagen-like triple helices (CO), and randomly distributed particles (PA) (Figure 1B). Section S1 of the supplementary document provides the details of the equations used for the generation of each design. It is important to note that the selected discretization strategy results in a significant discontinuity in the hard phase for the long OC design (*i.e.*, $W_G = 12$mm). We, nevertheless, included this design in the experimental groups to investigate the effects of such discontinuities on the mechanical performance of soft-hard interfaces. We included a control group without gradient transitions ($W_G = 0$ mm) (Figure S1A of the supplementary document). Furthermore, to measure the morphological features of each interface, we calculated the percentage of soft-hard normal contact area ($A_c$) across the gradient lengths (Figure S1B of the supplementary document). Finally, we projected each design into the narrow-gauge region of the tensile test specimens. Three specimens from each design were 3D printed, resulting in a total of 48 specimens.

## 2.3. Mechanical testing and post-processing

After manufacturing, we performed quasi-static uniaxial tensile tests with a mechanical testing bench (LLOYD instrument LR5K, load cell = 100 N) at a rate of 2 mm/min until failure. The device measured the displacements ($u$), forces ($f$), and time ($t$) at a sampling rate of 100 Hz. For all the tensile test specimens, we obtained full-field strain maps (the equivalent von Mises strains) at a frequency of 1 Hz using a 3D digital image correlation (DIC) system (Q-400, two cameras each with 12 MPixel, LIMESS GmbH, Krefeld, Germany) and its associated software (Instra 4D v4.6, Danted Dynamics A/S, Skovunde, Denmark). We, therefore, painted all the specimens white, followed by the application of a black dot speckle pattern.

To generate the stress-strain curves, we defined virtual extensometers at the center of the soft section of every specimen using the DIC software and extracted the vectors of true (logarithmic) strains ($\epsilon$). We then post-processed these vectors and their respective forces ($f$) in MATLAB R2018b (Mathworks, USA) to generate their true stress vectors ($\sigma = (f/A_o)\exp(\epsilon)$, $A_o = 32.512$ mm$^2$). From the resulting curves, we calculated the elastic modulus, $E$, as the slope of the linear region of the stress-strain curves measured between 0 and 20% strain, the ultimate tensile strength, $\sigma_{max}$, as the maximum recorded stress, and the strain energy density, $U_d$, as the area under the stress-strain curve, also known as toughness.

## 2.4. Finite element analysis of the interface designs

We created quasi-static finite element method (FEM) models of our designs using a commercially available software suite (nonlinear solver, Abaqus Standard v.6.13, Dassault Systèmes Simulia, France). Since the designs were symmetric across their length, we considered one longitudinal half of each design (Figure S2A of the supplementary document). Additionally, we only included one cross-sectional unit-cell of every design, yielding a perpendicular area of 24×24 elements. These simplifications led to discretized models with 110,592 hexagonal hybrid elements (C3D8H, enhanced hourglass control), where each voxel was represented as a single element. The hard phase was modeled as a linear elastic material with an elastic modulus $E$ of 2651 MPa and a Poisson's ratio, $\nu$, of 0.4, while the soft phase was modeled as an Ogden hyperelastic material with the following material parameters: $N = 1$, $\mu_1 = 0.266$ MPa, $\alpha_1 = 3.006, D_1 = 0.113$. Finally, we applied a surface traction of $\sigma_{FEM} = 0.186$ MPa to the hard end-surface and symmetric boundary conditions (*i.e.*, $U_x = R_y = R_z = 0$) to the soft end-surface of the mesh. The computationally predicted distributions of the von Mises strains were used for validation and analysis. Specifically, we compared the strain fields of the first transverse FEM layer with the DIC measurements.

## 2.5. Determination of maximum equivalent strains and strain concentration parameters

We used the true von Mises strains obtained from the DIC measurements and FEM simulations to study how the strains concentrated at the interfaces. To this end, we extracted the curves of the maximum equivalent strains $\epsilon_{eq,m}(x)$ from every cross-section layer across the length of the specimens. Additionally, we extracted the maximum strain value from these plots ($\epsilon_{max}$) and divided it over the average equivalent strain at the center of the specimen ($\epsilon_N$) to define the single-valued strain concentration parameter ($\epsilon_c = \epsilon_{max}/\epsilon_N$) (Figure 2A), which we obtained from the DIC measurements and FEM estimations. We used the strain values corresponding to a stress magnitude of $\sigma = 0.186$ MPa. Overall, these strain concentration parameters and the FEM-predicted strain distributions were used to study how the local geometry affected the performance of the functional gradients within the elastic loading regime.

**2.6. Determination of the elastic modulus functions**

We used the results from our computational models to study the elastic behavior exhibited by various designs. Towards this goal, we idealized the interfaces as linear systems of springs, where the elastic modulus function ($E(x) = \sigma_{FEM}/\epsilon_{avg}(x)$) is equivalent to the applied surface traction on the system, $\sigma_{FEM}$, over the average strain, $\epsilon_{avg}(x)$, of every transversal layer over each interface point (Figure S2B of the supplementary document). We used these estimated functions to study how different factors, including the elastic modulus and overall rigidity, affect the performance of the interfaces.

**2.7. Quad-lap shear designs, testing, post-processing, and finite element analysis**

We extended our analyses to study the shear response of the designed architectures using quad-lap shear test specimens. We chose the designs that performed the best in the tensile tests (*i.e.,* GY, CO, and PA, $W_G = 4.572$ mm) and compared them to the control design (*i.e.*, without a gradient). To obtain the appropriate dimensions of the specimens, we generated multiple geometries with the FEM software and simulated them. We used the dimensions of the specimen that yielded the elastic properties of the soft material ($E_{soft} = 1$ MPa, $v_{soft} = 0.49$) using only the forces and displacements of the virtual crosshead. After assigning the designs to the gradient regions, we printed these specimens (three per experimental group) and tested them under the same conditions as described above for the tensile tests. We calculated the true shear stresses ($\tau = f/(2tW), t = 3$ mm, $W = 33.528$ mm) and strains ($\gamma = d/(2H_S), H_S = 3.048$ mm) with the force and displacement vectors extracted from the mechanical testing machine. We then calculated the shear modulus ($G$, the initial slope of the stress-strain curves measured between 5% and 35% shear strain), maximum shear strength, $\tau_{max}$, and shear strain energy density, $U_d$, as the area under the shear stress-strain curve of each test. Furthermore, we performed FEM simulations of each design under similar conditions as described above for the modeling of quasi-static tensile tests. In this case, the unit cell used in the gradient region had a perpendicular area of 36×36 elements and a length of 144 elements (with no symmetry assumed), resulting in 186,624 hexagonal hybrid elements (C3D8H) per simulation. We assigned linear elastic (*i.e.*, $E = 2651$ MPa, $\nu = 0.4$) and hyperelastic (*i.e.*, Ogden, $N = 1$, $\mu_1 = 0.266$ Mpa, $\alpha_1 = 3.006, D_1 = 0.113$) properties to the hard and soft elements, respectively. Furthermore, we applied a shear deformation to one of the end-surfaces of the mesh (*i.e.*, $U_x = 0.4257$ mm, $R_x = R_y = R_z = 0$) while constraining all the displacements of the other. The results of the computational analysis were then used to calculate the strain concentration parameters of each design and for comparison with the experiments.

### 2.8. Design of a hybrid design based on our final findings

We further extended our analyses to exploit the full bioinspired potential of soft-hard interfaces. To do so, we combined multi-scale hierarchical organization, crack deflection mechanisms, functional gradients, and smooth contact areas to generate a hybrid design. We selected and combined the best-performing designs from the initial analysis, defining a PA structure in which the particles were randomly scattered within a GY architecture, resulting in the GP group. To create a less stiff elastic modulus function, we decreased the densities of the hard phase, $\rho$, by 50%. We then 3D printed, tested, and computationally analyzed these designs under the same conditions as described above for the quasi-static tensile tests.

### 3. RESULTS AND DISCUSSION

The integration of the different architectures into the soft-hard interfaces led to distinct patterns of $A_c$ (Figure 1B) and resulted in different total values of the contact surface area ($Tot.A_c$) (Figure 2B). None of these differences had a considerable effect on the initial elastic moduli (Figure 2C) calculated using the obtained stress-strain curves (Figure S3 of the supplementary document). Nevertheless, the varying geometries and gradient lengths did affect the strength and toughness of the interfaces (Figure 2D-E). The best-performing designs were GY ($W_G$ = 4 mm), CO ($W_G$ = 12 mm), and PA ($W_G$ = 12 mm). They all exhibited similar strengths and failure modes (*i.e.,* failure within the soft region), suggesting that the upper strength boundary of these interfaces was reached [34]. The control group under-performed all but the OC designs, confirming the importance of the implemented design strategies in improving the mechanical performance of soft-hard interfaces.

We found some evidence of the mechanism responsible for the failure of the control group specimens when assessing their DIC-measured EXP $\epsilon_c$ values (Figure 2F) and strain distributions (Figure 3A). According to this evidence, the shear strains at the edges of the interface were the primary culprits, which is consistent with the available literature [4,35,36]. These superficial strain concentrations were not present in the DIC results of any functionally graded design. Moreover, the EXP $\epsilon_c$ values of the other groups were all much lower than those measured in the control group. This lack of shear strains explains the improved performance of most of the presented designs, because a proper interfacing of soft and hard materials requires a smooth transition from one phase to another so that the stress concentration in the softer material can be decreased [37]. The FEM-predicted $\epsilon_c$ values (Figure 2G), however, indicated that strain concentrations take place within the 3D structure of the functionally graded interfaces. Therefore, a closer inspection to the results of the FEM simulations was necessary to elucidate the effects of the gradient morphology on the mechanical performance.

Comparing the DIC-measured strain distributions with the FEM results allowed us to validate our computational models (Figure 3B-F). In general, the predicted and measured strain distributions followed the same patterns. For example, all the GY designs showed curved-like strain patterns in both the experiments and simulations. In contrast, the strain distribution patterns of the DI and CO designs presented distinct diagonal lines. Similarly, the trends observed in the FEM $\epsilon_{eq,m}$ plots (of the first layer) resembled the DIC measurements. The simulations, however, showed higher peak strain values as compared to the experiments (*e.g.,* see the peaks at the edges of the DI designs or between individual voxels in the PA gradients). The absence of these peaks was likely caused by the limited DIC resolution (*i.e.,* between $22 \times 10^3$ and $27 \times 10^3$ facets per experiment), which was approximately six times lower than the resolution of the 3D-printed specimens (*i.e.,* $147 \times 10^3$ voxels). The FEM predictions were, therefore, more discerning when trying to understand the effects of geometrical design on the mechanical performance of soft-hard interfaces. This corroboration of the computational results allowed us to analyze the different designs individually in a quest to unravel the mechanisms underlying their mechanical performance.

Although generally better than the control group, most OC and DI specimens failed at the edge of the interface (only one DI specimen ($W_G = 8$ mm) failed at the center of the soft region). An analysis of the FEM-predicted strain distributions of these groups showed the prevalence of severe strain concentrations at their interface edges (Figure 3B-C), whose intensity was correlated to $W_G$. Upon closer inspection (Figure S4A-B of the supplementary document), the sharp-edged tips of the hard material seemed to have induced these strain concentrations. These shapes are particularly problematic since the interfacial geometry cannot arrest the propagation of initial cracks. Comparing the $A_c$ and $\epsilon_{eq,m}$ plots (Figure S5A-B of the supplementary document) of both designs indicated that these strain concentrations are associated with highly erratic $A_c$ patterns. A smooth material transition may, thus, alleviate such effects. It is, therefore, necessary to change the density of the hard phase, $\rho$, as smoothly as possible, use a geometry for which the change in $A_c$ is less abrupt, and ensure that there are no sharp ends in the selected geometrical design.

In the case of the long OC design (*i.e.,* $W_G = 12$ mm), the hard material discontinuity at the middle of the interface resulted in extreme strain concentrations (FEM $\epsilon_c = 3.38$) and was the region where critical cracks initiated. This lack of connectivity, which is visible in the discontinuous $A_c$ pattern of this design, led to extremely low values of interfacial strength and toughness (*i.e.,* approximately half the strength and toughness of the control group, Table S1 of the supplementary document). Therefore, although TPMS structures can yield closed-cell structures, verifying their connectivity by assessing their surface contact area and making the necessary corrections to $\rho$ is of great importance.

The GY results were particularly interesting because long gradients from this design ($W_G = 12$ mm) barely overperformed the control group, while the shorter version of the same design ($W_G = 4$ mm) outperformed all the other groups (Figure 2D-E). The short GY gradients presented failure modes where cracks initiated close to the interface but propagated through the soft region of the tensile specimens. In contrast, the other GY interfaces failed at the end of the interface. This performance difference can be due to several reasons. First, the GY specimens had $A_c$ patterns that were not as torturous as the OC and DI designs (Figure S5C of the supplementary document), explaining why their predicted FEM $\epsilon_c$ values were the lowest between the TPMS structures. More importantly, the strains of the short GY gradients mainly concentrated around the concave hard material shapes before the end of the gradient (Figure S4C of the supplementary document). This concave geometry, in turn, encased the regions with maximum strain concentrations, arresting the critical propagation of cracks. In comparison, the longer GY gradients showed tip-edged strain concentrations similar to those found in the OC and DI designs. Although the smooth $A_c$ pattern appears to have contributed to the high performance of short GY designs, comparison with the other designs indicates that the ability to contain the strain concentrations may have played a more important role in this regard, particularly given the fact that the short GY design was the only studied TPMS with this feature.

An important parameter affecting the performance of the soft-hard interface is the length over which the transition takes place (*i.e.*, $W_G$). The performance of a soft-hard interface is generally expected to improve as the length of the gradient increases, given that longer transitions lead to smoother changes in the elastic modulus, decreasing stress concentrations [37]. Indeed, the plots of the elastic modulus of the TPMS designs (Figure S4A-C of the supplementary document) were increasingly smoother as $W_G$ increased. The performance of the TPMS designs was, however, inversely related to $W_G$, as the strength values were higher for the specimens with shorter gradients while the FEM $\epsilon_c$ values were higher for the longer specimens. The local geometrical features at the end of a functional gradient are, therefore, more important in determining the mechanical performance of the interface than the overall smoothness of the function describing the transition of the elastic modulus. Consequently, it is important to utilize geometries that reduce the strain concentrations as $W_G$ increases or include features that help in arresting cracks.

In contrast with the TPMS structures, the mechanical performance of the CO and PA designs was enhanced as $W_G$ increased. In fact, the CO and PA designs with long gradients ($W_G = 12$ mm) were some of the toughest designs within this study. Moreover, most specimens of these two groups failed at the center of the soft region and not at the interface. When analyzing the computationally predicted strain distribution of the CO designs (Figure 3D, Figure S4D of the supplementary document), we observed that strains concentrated in the soft material regions between each coil of the functional gradient, resulting in smooth $\epsilon_{eq,m}$ plots (Figure S5D of the supplementary document) and the lowest FEM $\epsilon_c$ of this study. Such a proper distribution of strains may be attributed to the high $tot.A_c$ values and smooth $A_c$ patterns of these designs, which are similar to what is reported in the literature [35]. Furthermore, their elastic modulus functions were the most compliant, explaining the presence of strains across the longer sections of the gradient region (unlike in the TPMS results where the strains were concentrated at the edges). Particularly for the long CO design, the smooth and well-distributed strains across the entire gradient region indicated an increased strain energy storage capacity, leading to the high toughness values.

For the PA designs, the predicted FEM $\epsilon_c$ increased with $W_G$ but were not substantially higher than those of the control group. The locations of these strain concentrations were not necessarily at the end of the interface, but in single voxel locations across the entire gradient (Figure 3F, Figure S4E of the supplementary document). If a crack initiates around those stress concentration points, the nearby voxels could deflect it or arrest its progress, similar to what other studies have observed [38]. Furthermore, the comparatively higher magnitudes of strains across the gradient length produced more compliant elastic modulus functions in these designs, similar to the CO specimens (Figure S5E of the supplementary document). These high strains mean more strain energy is stored in such specimens, resulting in higher toughness values, particularly for the long PA gradients. Overall, both CO and PA achieved their high toughness because they could store more energy in their gradient region and because of the crack arresting features of their internal morphology.

We have so far only considered tension because this loading mode is typical in soft-hard interfaces (*e.g.,* cables, tendons, muscles). Soft-hard interfaces may, however, also fail under shear deformations, motivating the study of the presented designs under this loading regime. Since no standards are available for the geometrical design of functionally graded quad-lap specimens, a custom-made design was used (Figure 4A), enabling us to manufacture the specimens and perform mechanical testing (Figure 4B). After post-processing, all the hyperelastic $\tau - \gamma$ curves had similar initial values of the shear modulus ($G_{avg} = 0.308$ MPa, Figure 4C), confirming the proper design of the specimens and the satisfactory distribution of materials in the presented designs. In terms of the shear strength and toughness, however, the PA designs outperformed all other groups (Figure 4D). Moreover, the performance of the specimens was inversely related to their predicted FEM $\epsilon_c$, suggesting that the internal morphology of these interfaces was responsible for this outcome. Based on these observations, we concluded that using PA reinforcement in the design of soft-hard interfaces can enhance their performance under shear deformations.

In summary, the data and analysis presented above indicate that there are several morphological and mechanical principles that can lead to a tough soft-hard interface. First, the $A_c$ must be smooth to prevent any sudden changes in $\epsilon_{eq,m}$ across the functional gradient. Second, the corresponding $tot.A_c$ values should be as high as possible to decrease the overall magnitude of strain concentrations. Furthermore, it is important for the functional gradients to be increasingly more compliant, particularly at the edges of the interface. This compliance will lead to increased average deformations across the entire interface length. Higher amounts of strain energy can, therefore, be stored in the gradient region, leading to diffused stress concentrations and tougher soft-hard interfaces. Finally, the geometry of the structures should be capable of arresting cracks, particularly at the end of the gradient region. Examples of such geometries include concave designs implemented around strain concentration regions and randomly distributed particles. The selected geometries should also not include sharp tips of the hard phase at the edges of the interface because they create strain concentrations. Since most of the studied designs failed to implement all the aforementioned morphological features, we decided to extend our analysis by creating a design that combines the best performing TPMS (*i.e.,* the GY) with PA (Figure 5A-B). We hypothesized that adding particles to a gyroid design (GY + PA = GP) will hinder the propagation of critical cracks while producing a smooth $A_c$ pattern. Additionally, we decreased the ratio of the hard material, $\rho$, to 50% of its original value so that higher magnitudes of strain energy can be accommodated by the interface.

Our experiments confirmed that the GP specimens, indeed, exhibit many of the characteristics of high-performing soft-hard interfaces. The DIC-measured strain distributions of the GP specimens (Figure 5C) shared certain features with both the GY and PA designs while lacking any significant superficial strain concentrations. Overall, the 3D FEM models showed higher strain magnitudes and relatively higher levels of strain concentrations across the length of the functional gradient (Figure S4F of the supplementary document). All in all, the GP designs yielded the smoothest modulus functions in this study, enabling more strain energy to be stored within the FG. Furthermore, their $\epsilon_{eq,m}$ plots showed that the strain concentrations vanish before the end of the gradient region (Figure S5F of the supplementary document). Similar to the short GY and long PA designs, the presence of hard material around the locations of peak strains indicates that the interface can arrest initial cracks (Figure 5E). The performance of the GP specimens increased with $W_G$ (Figure 5D). In fact, the long ($W_G = 12$ mm) GP specimens presented the highest toughness values in this study (*i.e.,* 1.48 times tougher than the control specimens), confirming that the implementation of multi-scale features into a highly interdigitated and compliant functionally graded design further enhances the performance of soft-hard interfaces.

## 4. CONCLUSIONS

We studied how the design of soft-hard interfaces influences their mechanical performance. Our results clearly show the role of increased contact area, elastic modulus functions, and design features that attenuate or constrain strain concentrations in the rational design of high-performing soft-hard interfaces. The application of the abovementioned design features yielded soft-hard interfaces whose strength approached the upper boundary of the possible strengths and whose toughness increased by ≈ 50% as compared to that of the control specimens. Future work should employ these guidelines with computational methods to design optimized soft-hard interfaces. The presented results ultimately contribute to the development of the next generation of designer materials with applications in, among other areas, medical devices, tissue engineering, soft robotics, and the design of architected flexural mechanisms.

COMPETING INTERESTS

The authors declare no competing interests.

ACKNOWLEDGMENTS

This project is part of the Idea Generator (NWA-IDG) research program with code numbers NWA.1228.192.206 and NWA.1228.192.228. This research was conducted on a Stratasys® Objet350 Connex3™ printer through the Voxel Print Research Program. This program is an exclusive partnership with Stratasys Education that enhances the value of 3D printing as a powerful platform for experimentation, discovery, and innovation; for more information, contact: academic.research@stratasys.com.

SUPPLEMENTARY MATERIAL

Please see the supplementary file of this project.

REFERENCES

[1] V.L. Hein, F. Erdogan, Stress singularities in a two-material wedge, Undefined. 7 (1971) 317–330. https://doi.org/10.1007/BF00184307.

[2] R. Desmorat, F.A. Leckie, Singularities in bi-materials: parametric study of an isotropic/anisotropic joint, Eur. J. Mech. - A/Solids. 17 (1998) 33–52. https://doi.org/10.1016/S0997-7538(98)80062-4.

[3] Z. Wu, Stress concentration analyses of bi-material bonded joints without in-plane stress singularities, Int. J. Mech. Sci. 50 (2008) 641–648. https://doi.org/10.1016/J.IJMECSCI.2008.01.004.

[4] R.D. Adams, J. Coppendale, N.A. Peppiatt, Stress analysis of axisymmetric butt joints loaded in torsion and tension:, Http://Dx.Doi.Org/10.1243/03093247V131001. 13 (1978) 1–10. https://doi.org/10.1243/03093247V131001.

[5] R. Balokhonov, V. Romanova, ON THE PROBLEM OF STRAIN LOCALIZATION AND FRACTURE SITE PREDICTION IN MATERIALS WITH IRREGULAR GEOMETRY OF INTERFACES, Facta Univ. Ser. Mech. Eng. 17 (2019) 169–180. https://doi.org/10.22190/FUME190312023B.

[6] M.N. Saleh, M. Saeedifar, D. Zarouchas, S.T. De Freitas, Stress analysis of double-lap bi-material joints bonded with thick adhesive, Int. J. Adhes. Adhes. 97 (2020) 102480. https://doi.org/10.1016/J.IJADHADH.2019.102480.

[7] R. Lopes Fernandes, S. Teixeira de Freitas, M.K. Budzik, J.A. Poulis, R. Benedictus, Role of adherend material on the fracture of bi-material composite bonded joints, Compos. Struct. 252 (2020) 112643. https://doi.org/10.1016/J.COMPSTRUCT.2020.112643.

[8]  Z. Wu, A method for eliminating the effect of 3-D bi-material interface corner geometries on stress singularity, Eng. Fract. Mech. 73 (2006) 953–962. https://doi.org/10.1016/J.ENGFRACMECH.2005.10.010.

[9]  U.G.K. Wegst, H. Bai, E. Saiz, A.P. Tomsia, R.O. Ritchie, Bioinspired structural materials, Nat. Mater. 14 (2015) 23–36. https://doi.org/10.1038/nmat4089.

[10] F. Libonati, M.J. Buehler, Advanced Structural Materials by Bioinspiration , Adv. Eng. Mater. 19 (2017) 1600787. https://doi.org/10.1002/adem.201600787.

[11] J.W.C. Dunlop, P. Fratzl, Biological Composites, Annu. Rev. Mater. Res. 40 (2010) 1–24. https://doi.org/10.1146/annurev-matsci-070909-104421.

[12] A. Tits, D. Ruffoni, Joining soft tissues to bone: Insights from modeling and simulations, Bone Reports. 14 (2021) 100742. https://doi.org/10.1016/j.bonr.2020.100742.

[13] L. Rossetti, L.A. Kuntz, E. Kunold, J. Schock, K.W. Müller, H. Grabmayr, J. Stolberg-Stolberg, F. Pfeiffer, S.A. Sieber, R. Burgkart, A.R. Bausch, The microstructure and micromechanics of the tendon–bone insertion, Nat. Mater. 2017 166. 16 (2017) 664–670. https://doi.org/10.1038/nmat4863.

[14] C. Pitta Kruize, S. Panahkhahi, N.E. Putra, P. Diaz-Payno, G. Van Osch, A.A. Zadpoor, M.J. Mirzaali, Biomimetic Approaches for the Design and Fabrication of Bone-to-Soft Tissue Interfaces, ACS Biomater. Sci. Eng. (2021). https://doi.org/10.1021/ACSBIOMATERIALS.1C00620.

[15] S.P. Ho, S.J. Marshall, M.I. Ryder, G.W. Marshall, The tooth attachment mechanism defined by structure, chemical composition and mechanical properties of collagen fibers in the periodontium, Biomaterials. 28 (2007) 5238–5245. https://doi.org/10.1016/J.BIOMATERIALS.2007.08.031.

[16] A. Tits, E. Plougonven, S. Blouin, M.A. Hartmann, J.F. Kaux, P. Drion, J. Fernandez, G.H. van Lenthe, D. Ruffoni, Local anisotropy in mineralized fibrocartilage and subchondral bone beneath the tendon-bone interface, Sci. Reports 2021 111. 11 (2021) 1–17. https://doi.org/10.1038/s41598-021-95917-4.

[17] A.R. Studart, Biological and Bioinspired Composites with Spatially Tunable Heterogeneous Architectures, Adv. Funct. Mater. 23 (2013) 4423–4436. https://doi.org/10.1002/adfm.201300340.

[18] Z. Liu, M.A. Meyers, Z. Zhang, R.O. Ritchie, Functional gradients and heterogeneities in biological materials: Design principles, functions, and bioinspired applications, Elsevier Ltd, 2017. https://doi.org/10.1016/j.pmatsci.2017.04.013.

[19] A. Miserez, T. Schneberk, C. Sun, F.W. Zok, J.H. Waite, The Transition from Stiff to Compliant Materials in Squid Beaks, Science (80-. ). 319 (2008) 1816–1819. https://doi.org/10.1126/SCIENCE.1154117.

[20] M.E. Launey, M.J. Buehler, R.O. Ritchie, On the Mechanistic Origins of Toughness in Bone, Http://Dx.Doi.Org/10.1146/Annurev-Matsci-070909-104427. 40 (2010) 25–53. https://doi.org/10.1146/ANNUREV-MATSCI-070909-104427.

[21] H. Chai, J.J.-W. Lee, P.J. Constantino, P.W. Lucas, B.R. Lawn, Remarkable resilience of teeth, Proc. Natl. Acad. Sci. 106 (2009) 7289–7293.

https://doi.org/10.1073/PNAS.0902466106.

[22] E.L. Doubrovski, E.Y. Tsai, D. Dikovsky, J.M.P. Geraedts, H. Herr, N. Oxman, Voxel-based fabrication through material property mapping: A design method for bitmap printing, CAD Comput. Aided Des. 60 (2015) 3–13. https://doi.org/10.1016/j.cad.2014.05.010.

[23] C. Bader, D. Kolb, J.C. Weaver, S. Sharma, A. Hosny, J. Costa, N. Oxman, Making data matter: Voxel printing for the digital fabrication of data across scales and domains, Sci. Adv. 4 (2018). https://doi.org/10.1126/sciadv.aas8652.

[24] S. Hasanov, A. Gupta, A. Nasirov, I. Fidan, Mechanical characterization of functionally graded materials produced by the fused filament fabrication process, J. Manuf. Process. 58 (2020) 923–935. https://doi.org/10.1016/J.JMAPRO.2020.09.011.

[25] I. Maskery, L. Sturm, A.O. Aremu, A. Panesar, C.B. Williams, C.J. Tuck, R.D. Wildman, I.A. Ashcroft, R.J.M. Hague, Insights into the mechanical properties of several triply periodic minimal surface lattice structures made by polymer additive manufacturing, Polymer (Guildf). 152 (2018) 62–71. https://doi.org/10.1016/j.polymer.2017.11.049.

[26] O. Al-Ketan, R. Rezgui, R. Rowshan, H. Du, N.X. Fang, R.K.A. Al-Rub, Microarchitected Stretching-Dominated Mechanical Metamaterials with Minimal Surface Topologies, Adv. Eng. Mater. 20 (2018) 1800029. https://doi.org/10.1002/ADEM.201800029.

[27] O. Al-Ketan, R. Rowshan, R.K. Abu Al-Rub, Topology-mechanical property relationship of 3D printed strut, skeletal, and sheet based periodic metallic cellular materials, Addit. Manuf. 19 (2018) 167–183. https://doi.org/10.1016/J.ADDMA.2017.12.006.

[28] C.T. Thorpe, C. Klemt, G.P. Riley, H.L. Birch, P.D. Clegg, H.R.C. Screen, Helical sub-structures in energy-storing tendons provide a possible mechanism for efficient energy storage and return, Acta Biomater. 9 (2013) 7948–7956. https://doi.org/10.1016/J.ACTBIO.2013.05.004.

[29] X. Hu, P. Cebe, A.S. Weiss, F. Omenetto, D.L. Kaplan, Protein-based composite materials, Mater. Today. 15 (2012) 208–215. https://doi.org/10.1016/S1369-7021(12)70091-3.

[30] M.J. Mirzaali, A. Herranz de la Nava, D. Gunashekar, M. Nouri-Goushki, R.P.E. Veeger, Q. Grossman, L. Angeloni, M.K. Ghatkesar, L.E. Fratila-Apachitei, D. Ruffoni, E.L. Doubrovski, A.A. Zadpoor, Mechanics of bioinspired functionally graded soft-hard composites made by multi-material 3D printing, Compos. Struct. 237 (2020) 111867. https://doi.org/10.1016/J.COMPSTRUCT.2020.111867.

[31] M.J. Mirzaali, M. Cruz Saldívar, A. Herranz de la Nava, D. Gunashekar, M. Nouri-Goushki, E.L. Doubrovski, A.A. Zadpoor, Multi-Material 3D Printing of Functionally Graded Hierarchical Soft–Hard Composites, Adv. Eng. Mater. 22 (2020) 1901142. https://doi.org/10.1002/adem.201901142.

[32] W.D. Pilkey, D.F. Pilkey, Z. Bi, Peterson's Stress Concentration Factors, 4th Editio, John Wiley & Sons, 2020. https://doi.org/10.1002/9781119532552.

[33] ASTM D638, ASTM D638 - 14 Standard Test Method for Tensile Properties of Plastics,

ASTM Stand. (2004). https://www.astm.org/Standards/D638 (accessed September 28, 2021).

[34] T. Kuipers, R. Su, J. Wu, C.C.L. Wang, ITIL: Interlaced Topologically Interlocking Lattice for continuous dual-material extrusion, Addit. Manuf. 50 (2022) 102495. https://doi.org/10.1016/J.ADDMA.2021.102495.

[35] T.S. Lumpe, J. Mueller, K. Shea, Tensile properties of multi-material interfaces in 3D printed parts, Mater. Des. 162 (2019) 1–9. https://doi.org/10.1016/J.MATDES.2018.11.024.

[36] F. Liu, T. Li, X. Jiang, Z. Jia, Z. Xu, L. Wang, The effect of material mixing on interfacial stiffness and strength of multi-material additive manufacturing, Addit. Manuf. 36 (2020) 101502. https://doi.org/10.1016/J.ADDMA.2020.101502.

[37] J.H. Waite, H.C. Lichtenegger, G.D. Stucky, P. Hansma, Exploring Molecular and Mechanical Gradients in Structural Bioscaffolds†, Biochemistry. 43 (2004) 7653–7662. https://doi.org/10.1021/BI049380H.

[38] M.J. Mirzaali, M.E. Edens, A.H. de la Nava, S. Janbaz, P. Vena, E.L. Doubrovski, A.A. Zadpoor, Length-scale dependency of biomimetic hard-soft composites, Sci. Rep. 8 (2018) 12052. https://doi.org/10.1038/s41598-018-30012-9.

**FIGURE CAPTIONS**

**Figure 1.** A) The standard tensile test specimens furnished with a functional gradient connecting the hard and soft polymer phases (out-of-plane thickness = 4 mm). These designs were 3D printed using a polyjet multi-material 3D printer. B) All the initial designs and their calculated percentage of the soft-hard normal contact area ($A_c$). We combined three different values of the gradient length ($W_G$) with five different unit-cell geometries (*i.e.*, octo, diamonds, gyroids, collagen-like helices, and randomly distributed particles).

**Figure 2.** A) A representative example of a digital image correlation (DIC) measurement showing the regions from which the strain concentration parameters ($\epsilon_c$) were calculated (obtained at the point where the equivalent stresses were 0.1866 MPa). The bar plots represent the morphological and mechanical properties of the studied designs (mean ± SD). These include B) the total soft-hard normal contact area, C) measured elastic modulus, D) maximum strength, E) strain energy density, F) experimentally measured $\epsilon_c$, and G) FEM-predicted $\epsilon_c$.

**Figure 3**. Representative strain distributions measured using DIC and predicted using the FEM models as well as the maximum equivalent strain ($\epsilon_{eq,m}$) plots for every design. The study groups include the A) control group, B) octo, C) diamond, D) gyroid, E) collagen-like helices, and F) randomly distributed particles. The FEM strain distributions are presented for the first layer and at the location where the maximum strains were predicted, for validation and analysis purposes, respectively.

**Figure 4**. The specimens used in the shear tests and their corresponding results. A) The parametrized geometry of the quad-lap shear test specimens, with an out-of-plane thickness of 3 mm and $W_G$ = 3.048 mm. B) The selected geometries (*e.g.,* GY, CO, and PA), chosen due to their high tensile performance. The presented DIC measurements of the equivalent shear strains correspond to a shear stress of 0.523 MPa. C) The average (with shaded areas representing ± SD) shear stress *vs.* shear strain ($\tau$ *vs.* $\gamma$) curves for every design. D) The bar

plots represent the measured shear modulus ($G$), strain energy density ($U_d$), and estimated FEM $\epsilon_c$ for each design (mean ± SD).

**Figure 5.** Hybrid designs combining gyroid geometries with randomly distributed particles (GP) through a multi-scale approach. B) The magnitude of the $\rho$ functions were decreased to produce more compliant functional gradients while maintaining smooth $A_c$ patterns. C) Representative strain distributions measured using DIC and predicted using computational models as well as $\epsilon_{eq,m}$ plots for these designs. D) A scatterplot comparing the $\sigma_{max}$ vs. $U_d$ results for the GY, PA, and GP designs. The GP specimens were among the best-performing ones. E) A detailed comparison of the mechanistic features of the best-performing GY, PA, and GP designs.

**Figure 1**

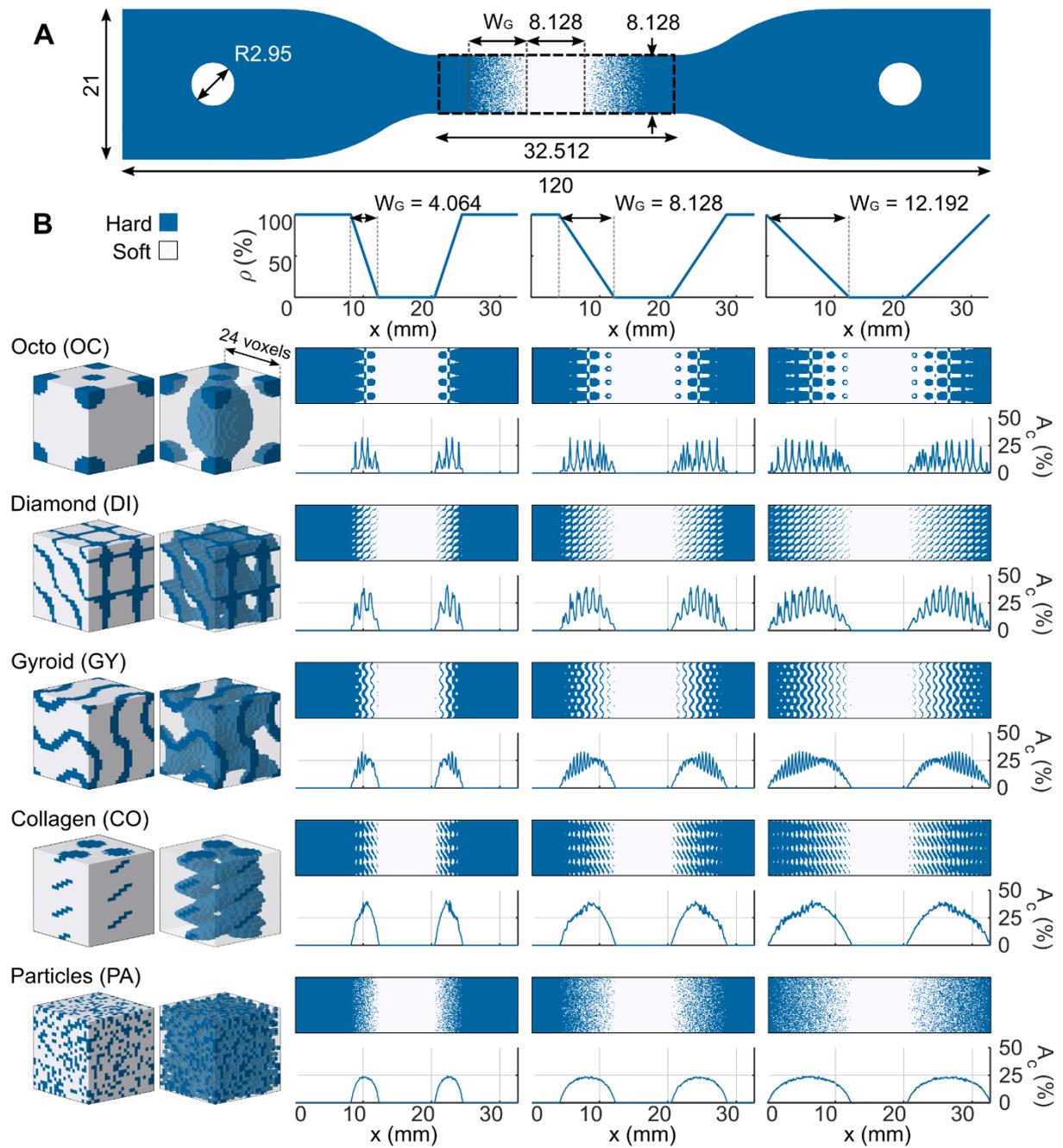

**Figure 2**

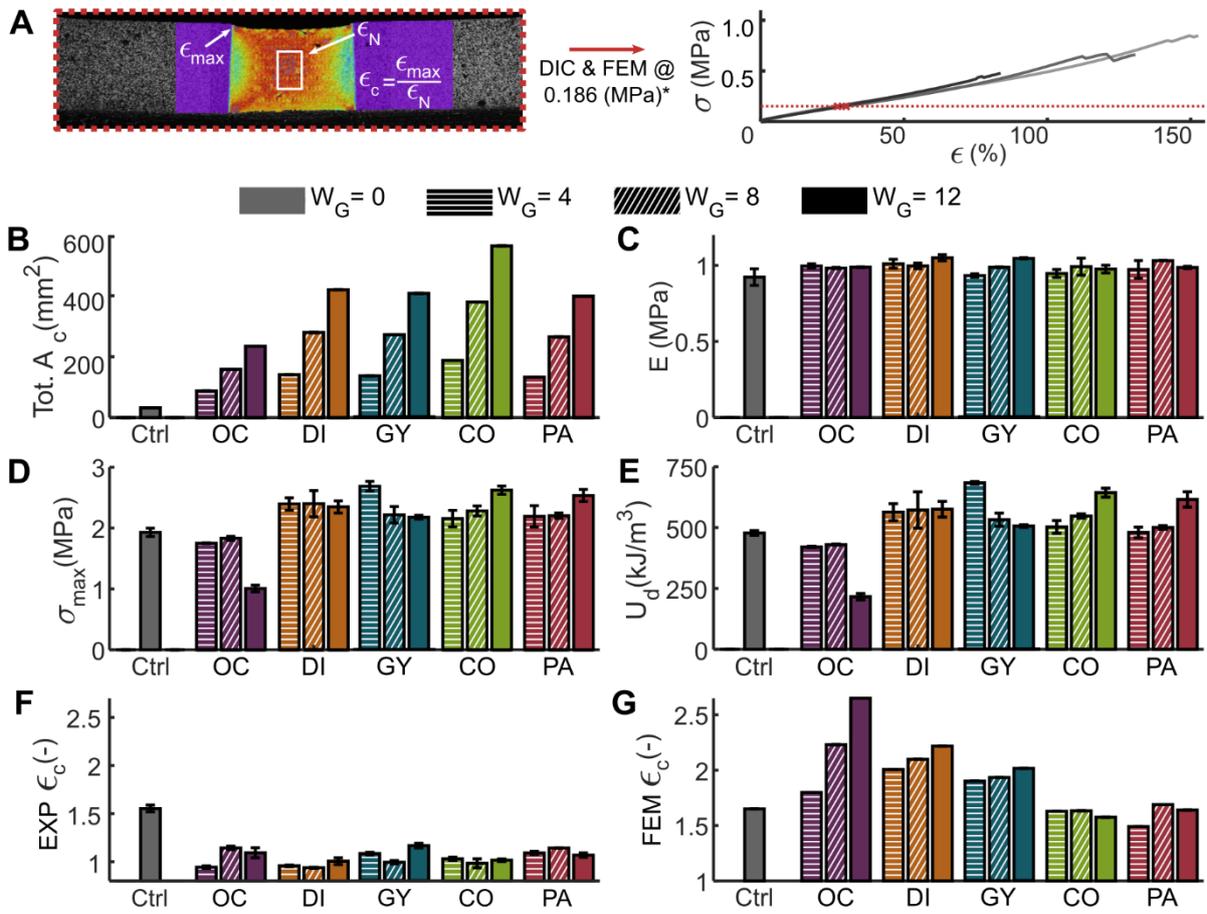

**Figure 3**

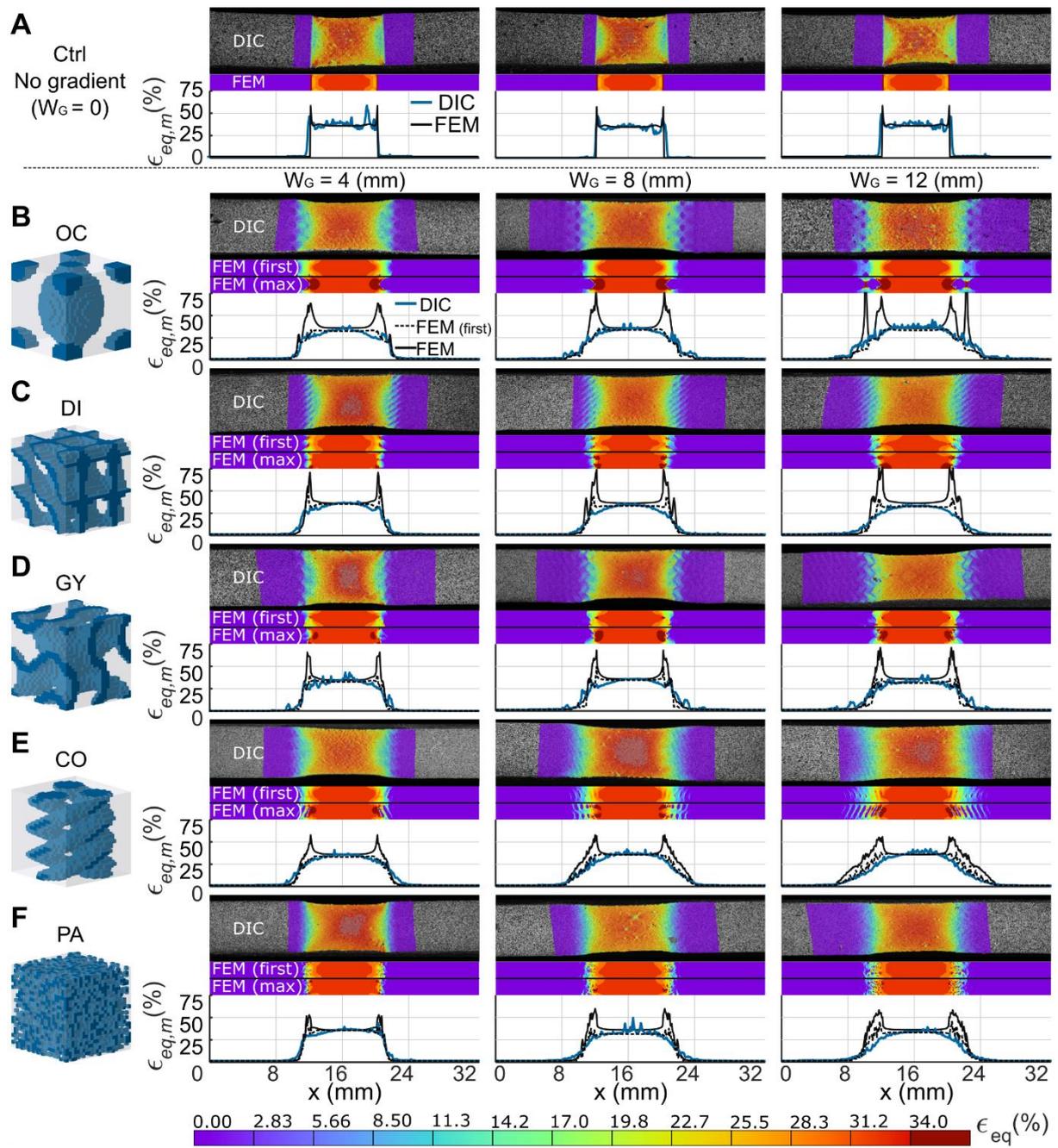

**Figure 4**

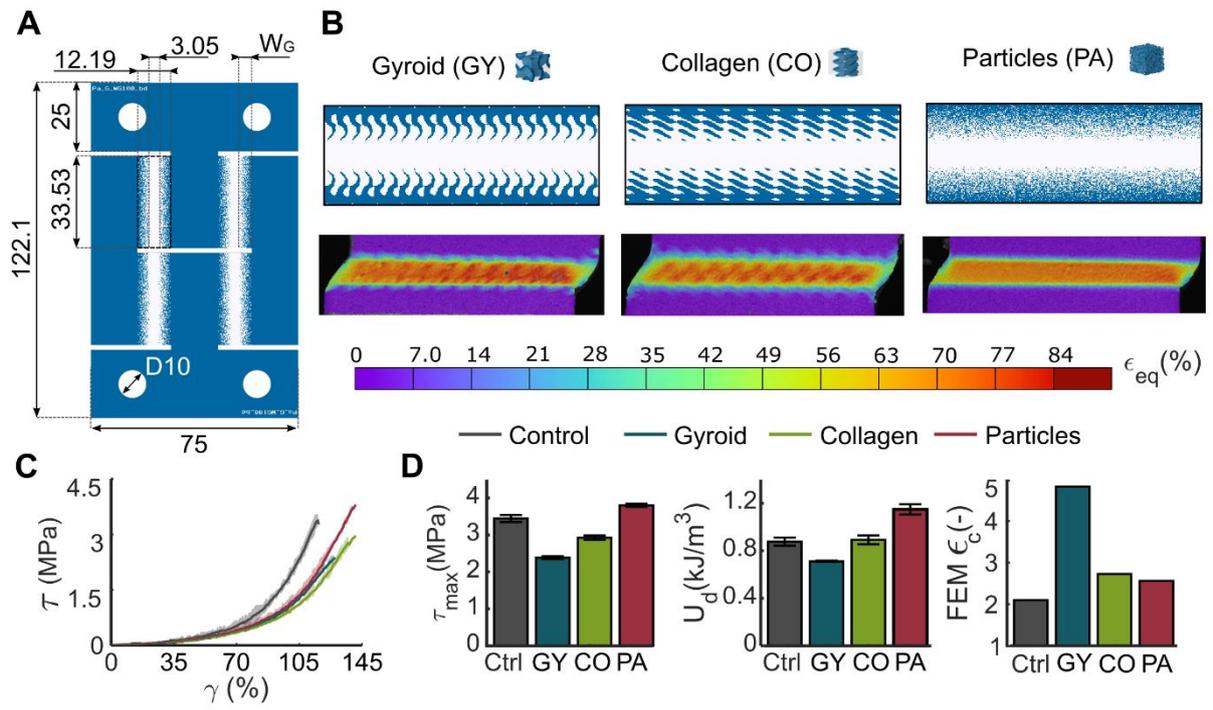

**Figure 5**

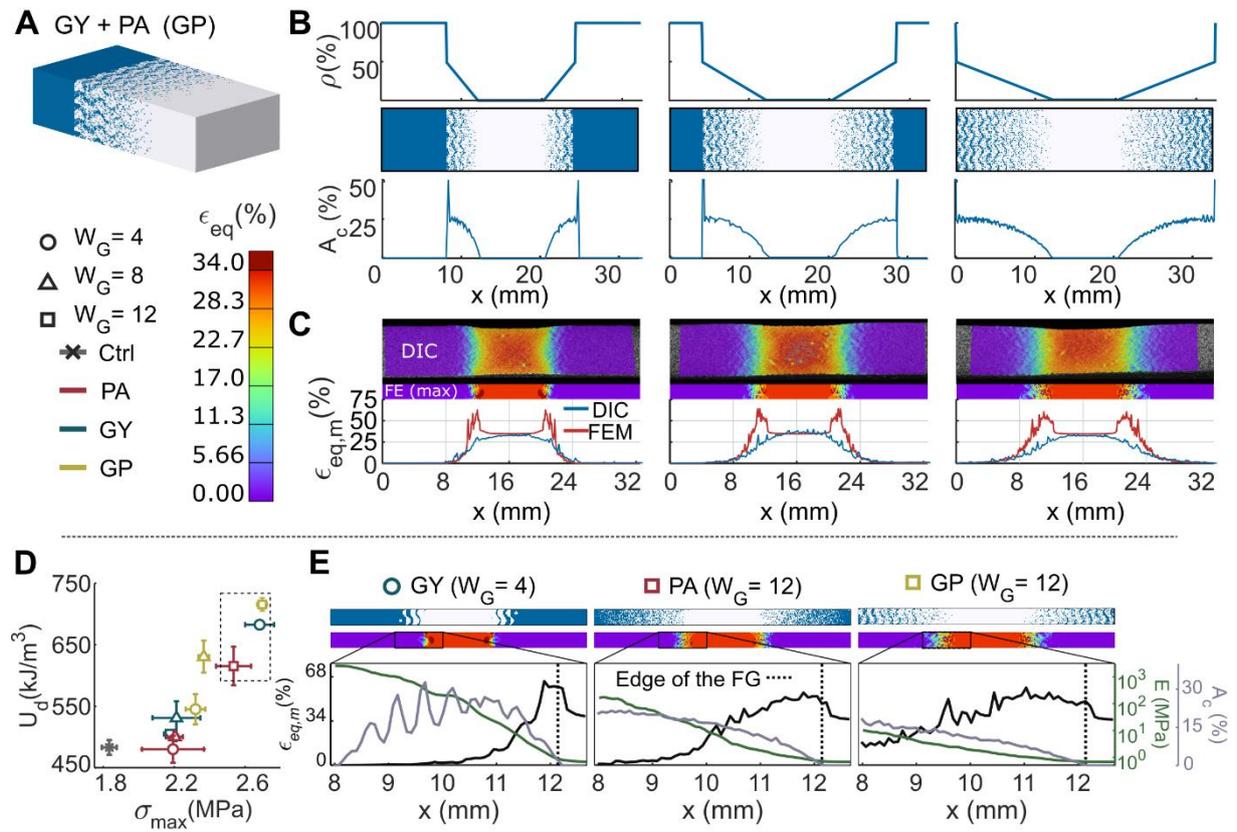